\documentstyle[12pt]{article}
\newlength{\extraspace}
\newlength{\extraspaces}
\newcommand{\be}{\begin{equation}
\addtolength{\abovedisplayskip}{\extraspaces}
\addtolength{\belowdisplayskip}{\extraspaces}
\addtolength{\abovedisplayshortskip}{\extraspace}
\addtolength{\belowdisplayshortskip}{\extraspace}}
\newcommand{\ee}{\end{equation}}

\newcommand{\ba}{\begin{eqnarray}
\addtolength{\abovedisplayskip}{\extraspaces}
\addtolength{\belowdisplayskip}{\extraspaces}
\addtolength{\abovedisplayshortskip}{\extraspace}
\addtolength{\belowdisplayshortskip}{\extraspace}}
\newcommand{\ea}{\end{eqnarray}}

\newcommand{\newsection}[1]{
\vspace{15mm}
\pagebreak[3]
\addtocounter{section}{1}
\setcounter{equation}{0}
\setcounter{subsection}{0}
\setcounter{footnote}{0}
\begin{flushleft}
{\large\bf \thesection. #1}
\end{flushleft}
\nopagebreak
\medskip
\nopagebreak}

\begin{document}

\addtolength{\baselineskip}{.8mm}

{\thispagestyle{empty}
\noindent \hspace{1cm}  \hfill July 1995 \hspace{1cm}\\
\mbox{}                 \hfill hep--th/9506043 \hspace{1cm}\\

\begin{center}\vspace*{1.0cm}
{\large\bf A remark on the high--energy quark--quark scattering}\\
{\large\bf and the eikonal approximation}\\
\vspace*{1.0cm}
{\large Enrico Meggiolaro}\\
\vspace*{0.5cm}{\normalsize
{Dipartimento di Fisica, \\
Universit\`a di Pisa, \\
I--56100 Pisa, Italy.}}\\
\vspace*{2cm}{\large \bf Abstract}
\end{center}

\noindent
In this paper we calculate the high--energy quark--quark scattering amplitude,
first in the case of scalar QCD, using Fradkin's approach to derive the
scalar quark propagator in an external gluon field and computing it in
the eikonal approximation.
(This approach was also recently used by Fabbrichesi, Pettorino, Veneziano
and Vilkovisky to study the four--dimensional
Planckian--energy scattering in gravity.)
We then extend the results to the case of
``real'' ({\it i.e.} fermion) QCD, thus deriving again, in a rather direct
way, the results previously found by Nachtmann.
The abelian case (QED) is also discussed in the Appendix.
}
\vfill\eject

\newsection{Introduction}

\noindent
High--energy scattering of strong--interacting particles attracted the
interest of many physicists between the late 1950s and the early 1960s,
when the first accelerators capable of producing protons in the $GeV$ region
were built. Before Quantum Chromo--Dynamics (QCD) was developed,
essentially two relevant models (both
originally based on potential scattering) were proposed to describe the
physics of strong--interacting particles: the {\it Regge pole model}
\cite{Regge} and
the {\it droplet model} of Yang and collaborators \cite{Yang}.
Although the physical
pictures of these two models were surely interesting and had some partial
experimental confirmations, they had not a strong theoretical
foundation upon first principles.

With the advent of quantum field theories, theoretical physicists became
more and more involved in the study of the high energy behaviour of gauge
field theories and in particular of Quantum Chromo--Dynamics,
which is now generally believed to
be the correct theory of hadrons and their interactions.
A lot of work has been done within the framework of perturbation theory in
order to find systematic procedures for extracting the large $s$, with
$t$ fixed, behaviour of each amplitude and for summing these
contributions using a leading--$\log$ or eikonal approximation scheme
\cite{Cheng-Wu-book} \cite{Lipatov}.

Anyway there is a class of {\it soft} high--energy scattering processes,
{\it i.e.} elastic scattering processes at high squared energies $s$ in
the center of mass and small squared transferred momentum $t$ (that is
$s \to \infty$ and $t \ll s$, let us say $|t| \le 1~GeV^2$), for which
QCD perturbation theory cannot be safely applied, since $t$ is too small.
Elaborate procedures for summing perturbative contributions have been
developed, even if the results are not able to explain the most relevant
phenomena.

A phenomenological analysis of this kind of {\it soft} hadronic reactions
at high energies
(in particular of the diffractive and elastic scattering of two hadrons)
can be performed using the so--called
{\it Pomeron--exchanged} model \cite{Pomeron}. Even if it is nowadays
accepted that the properties of the {\it Pomeron} come from the
multi--gluon exchange among the various partons in the process, it is still
an open problem how to quantitatively develop this picture in the
framework of Quantum Chromo--Dynamics.
Phenomenological analyses of the experimental results and a lot of theoretical
investigations suggest that the {\it Pomeron} couples to the single partons
within the hadrons, so that it is meaningful to study the parton--parton
scattering amplitude in detail, and in particular the quark--quark and
quark--antiquark amplitudes.

The first non--perturbative analysis, based on QCD, of these high--energy
scattering processes was performed by Nachtmann in \cite{Nachtmann91}.
He studied the $s$--dependence of the quark--quark (and quark--antiquark)
scattering amplitude by analytical means, using a functional integral
approach and an eikonal approximation to the solution of the Dirac equation
in the presence of a non--abelian external gluon field.

In this paper we propose a new approach to high--energy quark--quark (and
quark--antiquark) scattering, based on a first--quantized path--integral
description of quantum field theory developed by Fradkin in the early 1960s
\cite{Fradkin}. In this approach one obtains convenient expressions for the
full and truncated--connected scalar propagators in an external
(gravitational, electromagnetic, ...) field and the eikonal approximation
can be easily recovered in the relevant limit. Knowing the
truncated--connected propagators, one can then extract, {\it \`a la} LSZ,
the scattering matrix elements in the framework of a functional integral
approach. We remind that this same method has been recently adopted
in \cite{Veneziano} in order to study Planckian--energy
gravitational scattering. We shall make directly use of some results
contained in that paper, applying them to the case of interest to us,
{\it i.e.} Quantum Chromo--Dynamics.

This paper is organized as follows.
In Sect. 2 we calculate the high--energy quark--quark scattering amplitude
in the case of scalar QCD, using Fradkin's approach to derive the
scalar quark propagator in an external gluon field and then computing it in
the eikonal approximation. In Sect. 3 we extend the results to the case of
``real'' ({\it i.e.} fermion) QCD, thus deriving again, in a rather direct
way, the results previously found by Nachtmann in \cite{Nachtmann91}.
The high--energy quark--quark scattering amplitude turns out to be
described by the expectation value of two light--like Wilson lines, running
along the classical trajectories of the two colliding particles.

The abelian case is discussed in the Appendix, where we recover
the well--known result for the eikonal amplitude
of the high--energy scattering in QED \cite{Cheng-Wu}
\cite{Abarbanel-Itzykson} \cite{Jackiw}.

A discussion on the validity of the various approximations used
and on some calculation approaches proposed in the literature
is included and concludes the paper.

\vfill\eject

\newsection{The case of scalar QCD}

\noindent
We begin, for simplicity, with the case of {\it scalar} QCD, {\it i.e.}
the case of a spin--0 quark (described by the scalar field $\phi$) coupled
to a non--abelian gluon field $A^\mu \equiv A^\mu_a T_a$, $T_a$
($a=1, \ldots ,N_c^2 - 1$) being the generators of the Lie algebra of the
colour group $SU(N_c)$.
We limit ourselves to the case of one single flavour. The Lagrangian is:
\be
L(\phi ,\phi^\dagger ,A) =
[D^\mu \phi]^\dagger D_\mu \phi - m^2 \phi^\dagger \phi
- {1 \over 4} F^a_{\mu\nu} F^{a \mu\nu} ~,
\ee
where, as usual, $D^\mu = \partial^\mu + ig A^\mu$ is the covariant
derivative.
We then make use of Fradkin's approach \cite{Fradkin} to write the
propagator for the scalar field $\phi$ as a functional integral of the
first quantized theory \cite{Feynman}.
This approach was also recently used by Fabbrichesi, Pettorino, Veneziano
and Vilkovisky in \cite{Veneziano} to study the four--dimensional
Planckian--energy scattering in gravity. We shall use some results derived
in \cite{Veneziano}, adapting them to the case of interest to us,
{\it i.e.} high--energy scattering in scalar QCD. The extension to ``real'',
fermion, QCD will be discussed in the next section.

The Green function (Feynman propagator) for the scalar field $\phi$ in the
external metric $g_{\mu\nu}$ and in the external abelian field $A^\mu$ admits
a representation in terms of a functional integral over trajectories
\cite{Veneziano}:
\ba
\lefteqn{
G(y,x|g,A) = } \nonumber \\
& & = \displaystyle\int_0^\infty d\nu \int [\sqrt{-g(X)} dX^\mu]
\delta (x-X(0)) \delta (y-X(\nu)) \times \nonumber \\
& & \times \exp \left[ (i/2) \displaystyle\int_0^\nu d\tau
(g_{\mu\nu} \dot{X}^\mu
\dot{X}^\nu + 2e A_\mu \dot{X}^\mu -m^2) \right] ~.
\ea
As explained in ref.\cite{Bastianelli}, one can employ the path--integral
formulation of quantum mechanics also when $A^\mu$ is a matrix--valued
potential of the form $A^\mu \equiv A^\mu_a T_a$, where the matrices $T_a$
are the internal symmetry generators. As a consequence, the action $S$
itself is a matrix with internal symmetry indices and, in order to have
gauge covariance of the path--integral, it is necessary to define the
exponential of the action by path--ordering.
One can thus easily generalize the result (2.2) to the case of a
non--abelian external gluon field $A^\mu$:
\ba
\lefteqn{
G(y,x|g,A) = } \nonumber \\
& & = \displaystyle\int_0^\infty d\nu \int [\sqrt{-g(X)} dX^\mu]
\delta (x-X(0)) \delta (y-X(\nu)) \times \nonumber \\
& & \times {P} \exp \left[ (i/2) \displaystyle\int_0^\nu d\tau (g_{\mu\nu}
\dot{X}^\mu
\dot{X}^\nu \cdot {\bf 1} + 2g A_\mu \dot{X}^\mu -m^2 \cdot {\bf 1})
\right] ~,
\ea
where ${\bf 1}$ is the $N_c \times N_c$ unity matrix in the colour space
and ``$P$'' means ``{\it path--ordered}'' along the ``histories''
$X^\mu (\tau)$. Moreover $g(X) \equiv
\det (g_{\mu\nu} (X))$, so that $g(X) = -1$ in a flat space--time.
The propagator (2.3) is already connected, {\it i.e.} vacuum diagrams have
been already divided out. Yet, in order to derive the scattering matrix
elements following the LSZ approach, we need to know the on--shell
{\it truncated--connected} Green functions: these are obtained from the
connected Green functions by removing the external legs calculated
on--shell.
The truncated--connected propagator in the momentum space is defined as:
\be
\tilde{G}_t (p,p'|A) \equiv \mathop{\lim}_{p^2,p'^2 \to m^2}
{ p^2 - m^2 \over i} \tilde{G} (p,p'|A) { p'^2 - m^2 \over i} ~,
\ee
where $\tilde{G} (p,p'|A)$ is the Fourier transform of $G(x,y|A)$:
\be
\tilde{G} (p,p'|A) \equiv \int d^4 x \int d^4 y \exp [i(p'x - py)]
G(x,y|A) ~,
\ee
with: $G(x,y|A) \equiv <T(\phi(x) \phi^\dagger (y) >_A$.
In ref.\cite{Veneziano} it is shown how to evaluate the
truncated--connected propagator (2.4) starting from eq.(2.3).
It is also shown how to compute $\tilde{G}_t (p,p'|A)$ in the so--called
{\it eikonal} approximation, which is valid in the case of scattering
particles with very high energy ($E \equiv p^0 \simeq |{\bf p}| \gg m$) and
small transferred momentum $q \equiv p' - p$ ({\it i.e.} $|t| \ll E$, where
$t = q^2$). In this limit the functional integral can be evaluated by means
of a saddle--point approximation in which the classical trajectory is
computed to the lowest non--trivial order, that is a straight line (in the
Minkowski space--time) described by a classical particle (of mass $m$)
moving with 4--momentum $p^\mu$:
\be
X^\mu_0 (\nu) = b^\mu + {p^\mu \over m} \nu ~.
\ee
One thus finds, after properly generalizing the results of \cite{Veneziano}
to the case of a non--abelian gluon external field, the following
expression for the truncated--connected propagator in the eikonal
approximation:
\be
\tilde{G}_t (p,p'|A) \simeq 2E \displaystyle\int d^3 b e^{iqb}
\left( {P} \exp \left[ -ig \displaystyle\int_{-\infty}^{+\infty}
A_\mu (b + p\tau) p^\mu d\tau \right] - {\bf 1} \right) ~.
\ee
As said before, $q = p' - p$ is the transferred momentum. In ``$d^3 b$''
one must not include the component of $b^\mu$ which is parallel to $p^\mu$.
For example, if $p^\mu \simeq p'^\mu \simeq (E,E,0,0)$, one has that
$p_+ \simeq p'_+ \simeq 2E$ and $p_- \simeq p'_- \simeq 0$, where the
following general notation has been used for a given 4--vector $A^\mu$:
\be
A_+ \equiv A^0 + A^1 ~~~ , ~~~ A_- \equiv A^0 - A^1 ~.
\ee
In this case one must take: $d^3 b = d^2 {\bf b}_t db_-$, where
${\bf b}_t = (b_y, b_z)$ is the component of $b^\mu$ in the {\it
transverse} plane $(y,z)$ (while the two light--cone coordinates,
$x_+ = t + x$ and $x_- = t - x$, are sometimes
called ``{\it longitudinal}'' coordinates).

The expression (2.7) for the truncated--connected propagator in the
eikonal approximation will be our starting point for deriving the scattering
amplitude of two high--energy spin--0 quarks, in the limit $s \to \infty$
and $t \ll s$. We thus consider the elastic scattering process of two (scalar)
quarks with initial 4--momenta $p_1$ and $p_2$ and final 4--momenta $p'_1$
and $p'_2$:
\be
\phi(p_1) \phi(p_2) \rightarrow \phi(p'_1) \phi(p'_2) ~.
\ee
In the Center--of--Mass reference System (CMS), taking the initial
trajectories of the two quarks along the $x$--axis, the 4--momenta
$p_1$, $p_2$, $p'_1$ and $p'_2$ are given, in the first approximation, by:
\be
p_1 \simeq p'_1 \simeq (E,E,{\bf 0}_t) ~~~ , ~~~
p_2 \simeq p'_2 \simeq (E,-E,{\bf 0}_t) ~.
\ee
Using the LSZ reduction formula and a functional integral approach, one
finds the following expression for the scattering matrix element (with the
plane wave functions normalized as: $\phi_p (x) = \exp(-ipx)$):
\ba
\lefteqn{
<\phi(p'_1) \phi(p'_2) | ( S - {\bf 1} ) | \phi(p_1) \phi(p_2) > =}
\nonumber \\
& & = {1 \over Z_\phi^2} \{ < \tilde{G}_t (p_1,p'_1|A) \tilde{G}_t (p_2,
p'_2|A) >_A + < \tilde{G}_t (p_1,p'_2|A) \tilde{G}_t (p_2,
p'_1|A) >_A \} ~.
\ea
$Z_\phi$ is the renormalization constant of the field $\phi$:
$\phi_R (x) = Z_\phi^{-1/2} \phi (x)$.
The expectation value $< O (A) >_A$ of an arbitrary functional $O (A)$ of
the gluon field $A^\mu$ is defined as:
\be
< O (A) >_A \equiv {1 \over Z} \displaystyle\int [dA] O (A)
\exp \left[ -{i \over 4} \displaystyle\int d^4 x F_a^{\mu\nu} F_{a \mu\nu}
\right] \{ \det [ D^\mu D_\mu + m^2 ] \}^{-1} ~,
\ee
where $Z$ is the partition function for scalar QCD:
\ba
\lefteqn{
Z \equiv \displaystyle\int [dA] [d\phi] [d\phi^\dagger]
\exp \left[ i \int d^4 x L (\phi ,\phi^\dagger ,A) \right] }
\nonumber \\
& & = \displaystyle\int [dA]
\exp \left[ -{i \over 4} \displaystyle\int d^4 x F_a^{\mu\nu} F_{a \mu\nu}
\right] \{ \det [ D^\mu D_\mu + m^2 ] \}^{-1} ~.
\ea
The determinant in eqs.(2.12) and (2.13) comes from the integration over the
scalar degrees of freedom. In fact, the Lagrangian $L (\phi ,\phi^\dagger ,
A)$ in (2.1) is bilinear in the scalar fields $\phi$ and $\phi^\dagger$:
\be
\displaystyle\int d^4 x \left[ (D^\mu \phi)^\dagger D_\mu \phi - m^2
\phi^\dagger \phi \right] =
- \displaystyle\int d^4 x \phi^\dagger \left[ D^\mu D_\mu + m^2 \right]
\phi ~.
\ee
Therefore the functional integral over the scalar fields $\phi$ and
$\phi^\dagger$ is an ordinary Gaussian integral and can be
performed in the standard way, to give (apart from an irrelevant constant)
$\{ \det [ D^\mu D_\mu + m^2 ] \}^{-1}$.

The first expectation value in eq.(2.11) corresponds to the $t$--channel
scattering of
the two particles, with squared transferred momentum equal to:
$(p_1 - p'_1)^2 \equiv t \ll s$.
The second expectation value corresponds instead to the $u$--channel
scattering of the two quarks. In other words, the squared transferred
momentum, flowing from one quark to the other, is equal to:
\be
(p_1 - p'_2)^2 \equiv u = 4 m^2 -s -t \simeq -s ~,
\ee
in the limit we are considering: $s \to \infty$ with $t,~m^2 \ll s$.
In this high--energy limit the contribution coming from the second
expectation value in eq.(2.11) is smaller by at least a factor of $s$, when
compared with the first expectation value, and hence is negligible.
One can be easily convinced of this by considering the Feynman diagrams of
the process in the perturbation theory: the diagrams which correspond to the
second piece in eq.(2.11) have intermediate gluons carrying a big
squared transferred momentum $u \simeq -s$, so that their propagators
suppress the corresponding amplitude.
Therefore, in our limit:
\be
<\phi(p'_1) \phi(p'_2) | ( S - {\bf 1} ) | \phi(p_1) \phi(p_2) >
\simeq
{1 \over Z_\phi^2} < \tilde{G}_t (p_1,p'_1|A) \tilde{G}_t (p_2,
p'_2|A) >_A ~.
\ee
The diffusion amplitude $M_{fi} = < f | M | i >$ is given, as usual,
by:
\ba
\lefteqn{
<\phi(p'_1) \phi(p'_2) | ( S - {\bf 1} ) | \phi(p_1) \phi(p_2) > = }
\nonumber \\
& & = i (2\pi)^4 \delta^{(4)} (P_{fin} - P_{in})
<\phi(p'_1) \phi(p'_2) | M | \phi(p_1) \phi(p_2) > ~,
\ea
where $P_{in} = p_1 + p_2$ is the total initial 4--momentum and
$P_{fin} = p'_1 + p'_2$ is the total final 4--momentum.
In the following we shall evaluate the expectation value:
\be
\tilde{f} (s,t) \equiv
< \tilde{G}_t (p_1,p'_1|A) \tilde{G}_t (p_2,p'_2|A) >_A ~.
\ee
Let us indicate with $x_1^\mu (\tau)$ and $x_2^\mu (\tau)$ the classical
trajectories of the two colliding particles in the Minkowski space--time:
\be
x_1^\mu (\tau) = b_1^\mu + p_1^\mu \tau ~~~ , ~~~
x_2^\mu (\tau) = b_2^\mu + p_2^\mu \tau ~.
\ee
By virtue of eq.(2.10) we can thus write the line--integral in the eikonal
propagator for the particle 1 as:
\be
\displaystyle\int_{-\infty}^{+\infty} A_\mu (x_1) dx_1^\mu \simeq
{1 \over 2} \displaystyle\int_{-\infty}^{+\infty} A_- (x_1) dx_{1+} =
{1 \over 2} \displaystyle\int_{-\infty}^{+\infty} A_- (x_{1+},b_{1-},
{\bf b}_{1t}) dx_{1+} ~.
\ee
Similarly, for the particle 2:
\be
\displaystyle\int_{-\infty}^{+\infty} A_\mu (x_2) dx_2^\mu \simeq
{1 \over 2} \displaystyle\int_{-\infty}^{+\infty} A_+ (x_2) dx_{2-} =
{1 \over 2} \displaystyle\int_{-\infty}^{+\infty} A_+ (b_{2+},x_{2-},
{\bf b}_{2t}) dx_{2-} ~.
\ee
Therefore, if we define the two Wilson paths:
\ba
W_1(b_1) = W_1(b_{1+}, b_{1-}, {\bf b}_{1t}) &=&
{P} \exp \left[ -{i \over 2} g
\displaystyle\int_{-\infty}^{b_{1+}} A_- (x_{1+},b_{1-},
{\bf b}_{1t}) dx_{1+} \right] , \nonumber \\
W_2(b_2) = W_2(b_{2+}, b_{2-}, {\bf b}_{2t}) &=&
{P} \exp \left[ -{i \over 2} g
\displaystyle\int_{-\infty}^{b_{2-}} A_+ (b_{2+},x_{2-},
{\bf b}_{2t}) dx_{2-} \right] ~,
\ea
we obtain the following expression for the eikonal propagator (2.7)
of the particle 1:
\ba
\lefteqn{
\tilde{G}_t (p_1,p'_1|A) \simeq }
\nonumber \\
& & \simeq 2E \displaystyle\int d^2 {\bf b}_{1t} db_{1-} e^{i q_1 b_1}
\displaystyle\int_{-\infty}^{+\infty} db_{1+}
{\partial \over \partial b_{1+}} W_1 (b_{1+},b_{1-},{\bf b}_{1t})
\nonumber \\
& & \simeq 4E \displaystyle\int d^4 b_1 e^{iq_1 b_1}
{\partial \over \partial b_{1+}} W_1 (b_1) ~.
\ea
We have used the fact that $W_1 (-\infty ,b_{1-},{\bf b}_{1t}) = {\bf 1}$;
moreover $q_{1-} b_{1+} \simeq 0$, since $p_{1-} \simeq p'_{1-} \simeq 0$.
We proceed in a similar way for the eikonal propagator of the particle 2,
using the fact that $W_2 (b_{2+},-\infty ,{\bf b}_{2t}) = {\bf 1}$ and
$q_{2+}b_{2-} \simeq 0$ (since $p_{2+} \simeq p'_{2+} \simeq 0$). We thus
obtain that:
\be
\tilde{G}_t (p_2, p'_2|A) \simeq
4E \displaystyle\int d^4 b_2 e^{iq_2 b_2}
{\partial \over \partial b_{2-}} W_2 (b_2) ~.
\ee
We can substitute eqs.(2.23) and (2.24) into eq.(2.18), thus obtaining the
following expression for $\tilde{f} (s,t)$:
\ba
\lefteqn{
\tilde{f} (s,t) \equiv
< \tilde{G}_t (p_1,p'_1|A) \tilde{G}_t (p_2,p'_2|A) >_A }
\nonumber \\
& & \simeq 16 E^2
\displaystyle\int d^4 b_1 \displaystyle\int d^4 b_2
e^{i(q_1 b_1 + q_2 b_2)}
{\partial \over \partial b_{1+}}{\partial \over \partial b_{2-}}
< W_1 (b_1 - b_2) W_2 (0) >_A ~.
\ea
We have made use of the translational invariance of the expectation value
$< W_1 (b_1)$ $W_2 (b_2) >_A$ to write it as $< W_1 (b_1 - b_2) W_2 (0) >_A$.
In order to isolate, inside $\tilde{f} (s,t)$, the 4--dimensional
{\it delta}--function coming from the
conservation of the total 4--momentum (which we expect on the basis of
eq.(2.17)), it will be helpful to define the new variables $\Sigma$ and $z$
in the CMS:
\be
\Sigma = {b_1 + b_2 \over 2} ~~~ , ~~~ z = b_1 - b_2 ~.
\ee
It is clear that: $d^4 b_1 d^4 b_2 = d^4 \Sigma d^4 z$. Moreover:
\be
{\partial  \over \partial b_{1+}} = {1 \over 2}
{\partial  \over \partial \Sigma_+} + {\partial  \over \partial z_+} ~~~ ,
{}~~~ {\partial  \over \partial b_{2-}} = {1 \over 2}
{\partial  \over \partial \Sigma_-} - {\partial  \over \partial z_-} ~.
\ee
Still making use of the translational invariance of the expectation value
$< W_1 (b_1 - b_2) W_2 (0) >_A = < W_1 (z) W_2 (0) >_A$, one can write, after
some trivial algebra:
\ba
\lefteqn{
\tilde{f} (s,t) = 16 E^2
\displaystyle\int d^4 \Sigma e^{i(q_1 + q_2)\Sigma}
\displaystyle\int d^4 z e^{{i \over 2}(q_1 - q_2)z}
\times } \nonumber \\
& & \times
\left( -{\partial \over \partial z_+}{\partial \over \partial z_-} \right)
< W_1 (z_+, 0, {\bf z}_t) W_2 (0, -z_-, {\bf 0}_t) >_A
\nonumber \\
& & = (2\pi)^4 \delta^{(4)} (q_1 + q_2) \cdot 8E^2
\displaystyle\int d^2 {\bf z}_t \displaystyle\int dz_+ \displaystyle\int
dz_- e^{iq_1 z}
\times \nonumber \\
& & \times
\left( -{\partial \over \partial z_+}{\partial \over \partial z_-} \right)
< W_1 (z_+, 0, {\bf z}_t) W_2 (0, -z_-, {\bf 0}_t) >_A ~.
\ea
Since $q_{1-} \simeq 0$ ($p_{1-} \simeq p'_{1-} \simeq 0$) and
$q_{2+} \simeq 0$ ($p_{2+} \simeq p'_{2+} \simeq 0$), from the
{\it delta}--function in front of eq.(2.28) one also derives that:
$q_{1+} = -q_{2+} \simeq 0$ and $q_{2-} = -q_{1-} \simeq 0$.
Therefore one can take the phase--factor $\exp(iq_1 z)$ out of the
integrals in the light--cone variables $z_+$ and $z_-$.
Thus, defining:
\be
\tilde{f} (s,t) = i (2\pi)^4 \delta^{(4)} (P_{fin} - P_{in}) f (s,t) ~,
\ee
we finally find that (let us observe that $s \simeq 4E^2$ and
$q_1 + q_2 = (p'_1 + p'_2) - (p_1 + p_2) = P_{fin} - P_{in}$):
\be
f(s,t) = -i 2s \displaystyle\int d^2 {\bf z}_t e^{i {\bf q} \cdot {\bf z}_t}
< [ W_1 (+\infty , 0, {\bf z}_t) - {\bf 1} ] \times
[ W_2 (0, +\infty , {\bf 0}_t) - {\bf 1} ] >_A ~.
\ee
In the last equation we have put:
\be
q \equiv p_1 - p'_1 = -q_1 \simeq (0,0, -{\bf q}_{1t})
\equiv (0,0, {\bf q}) ~.
\ee
That is: ${\bf q} \equiv -{\bf q}_{1t}$. So that:
$q z \simeq -{\bf q} \cdot {\bf z}_t$. The expression (2.30) gives the
amplitude as an explicit function of the squared transferred momentum
$t = q^2 = -{\bf q}^2$ (therefore: $t < 0$ !). More explicitly, putting in
evidence the colour indices of the scalar quarks, the scattering amplitude
can be written as:
\ba
\lefteqn{
M_{fi} = < \phi_i (p'_1) \phi_k (p'_2) | M | \phi_j (p_1) \phi_l (p_2) >
= {1 \over Z_\phi^2} f(s,t) } \nonumber \\
& & \mathop{\simeq}_{s \to \infty} -{i \over Z_\phi^2} 2s
\displaystyle\int d^2 {\bf z}_t e^{i {\bf q} \cdot {\bf z}_t}
< [ W_1 (z_t) - {\bf 1} ]_{ij} [ W_2 (0) - {\bf 1} ]_{kl} >_A ~.
\ea
Therefore, the high--energy ($s \to \infty$ and $t \ll s$) scalar
quark--quark scattering amplitude turns out to be the Fourier transform,
with respect to the transverse coordinates ${\bf z}_t$,
of the expectation value of two
light--like Wilson lines, separated by $z_t = (0,0,{\bf z}_t)$:
\ba
W_1 (z_t) \equiv W_1 (+\infty , 0, {\bf z}_t) &=&
{P} \exp \left[ -ig \displaystyle\int_{-\infty}^{+\infty}
A_\mu (z_t + p_1 \tau) p_1^\mu d\tau \right] ~;
\nonumber \\
W_2 (0) \equiv W_2 (0, +\infty , {\bf 0}_t) &=&
{P} \exp \left[ -ig \displaystyle\int_{-\infty}^{+\infty}
A_\mu (p_2 \tau) p_2^\mu d\tau \right] ~.
\ea
The space--time configuration of these two Wilson lines is shown in [Fig.1].
This picture was originally proposed by Nachtmann in ref.\cite{Nachtmann91}
(for the case of ``real'' fermion QCD, which we shall
discuss in the next section).
He derived the $s$--dependence of the quark--quark (or quark--antiquark)
scattering amplitude using a functional integral approach and an eikonal
approximation to the solution of the Dirac equation
in the presence of a non--abelian external gluon field.

\newcommand{\wilson}[1]{
\begin{figure}
\begin{center}
\setlength{\unitlength}{1.00mm}
\raisebox{-40\unitlength}
{\mbox{\begin{picture}(80,45)(-35,-30)
\thicklines
\put(-22,22){\line(1,-1){41}}
\put(-15,15){\vector(-1,1){1}}
\put(16,23){\line(-1,-1){42}}
\put(9,16){\vector(1,1){1}}
\put(0,0){\vector(-2,-1){13}}
\thinlines
\put(-8,0){\line(1,0){35}}
\put(8,4){\line(-2,-1){35}}
\put(0,-8){\line(0,1){35}}
\put(27,0){\vector(1,0){1}}
\put(0,27){\vector(0,1){1}}
\put(-13,20){\makebox(0,0){$W_2$}}
\put(18,20){\makebox(0,0){$W_1$}}
\put(-6,-7){\makebox(0,0){$z_t$}}
\put(25,-2){\makebox(0,0){$x$}}
\put(-2,25){\makebox(0,0){$t$}}
\end{picture}}}
\parbox{13cm}{\small #1}
\end{center}
\end{figure}}

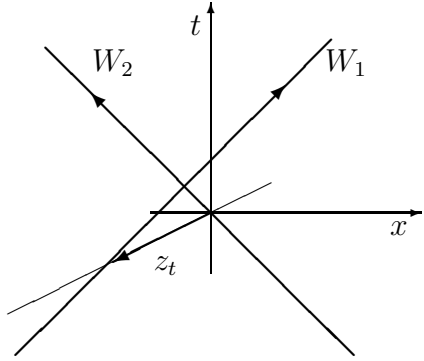
\wilson{{\bf Fig.~1.} The space--time configuration of the two light--like
Wilson lines $W_1$ and $W_2$ entering in the expression (2.32) for the
high--energy quark--quark elastic scattering amplitude.}

{}From eq.(2.32) it seems that the
$s$--dependence of the scattering amplitude is all contained in the
kinematic factor $2s$ in front of the integral. Yet, as it was pointed
out by H. Verlinde and E. Verlinde in \cite{Verlinde}, this is not true:
in fact, one can easily be convinced (for example by making a perturbative
expansion) that it is a singular limit to take the Wilson lines in
(2.32) exactly light--like. As suggested in \cite{Verlinde}, one can
regularize this sort of ``infrared'' divergence by letting each line to
have a small time--like component, so that they coincide with the classical
trajectories for quarks with finite mass $m$. In this way one obtains
a $\log s$--dependence of the amplitude, as expected from ordinary
perturbation theory. We will return to comment on this at the end of the
next section.

\newsection{The case of ``real'' (fermion) QCD}

\noindent
We want now to extend the results obtained in the previous section to the
case (more interesting from the physical point of view) of ``real''
fermion QCD: that is a non--abelian gluon field coupled to a
spin--${1 \over 2}$ quark.

As before, we limit ourselves to the case of one single flavour.
The QCD Lagrangian is:
\be
L(\psi ,\psi^\dagger ,A) =
\bar{\psi} (i\gamma^\mu D_\mu - m ) \psi
- {1 \over 4} F^a_{\mu\nu} F^{a \mu\nu} ~,
\ee
where, as usual, $D^\mu = \partial^\mu + ig A^\mu$ is the covariant
derivative.

Proceeding exactly as in the previous section, we
use the LSZ reduction formula and a functional integral approach to write
the quark--quark scattering matrix element (with the
plane wave functions normalized as: $\psi_p (x) = u_p \exp(-ipx)$, with
$\bar{u}_p u_p = 2m$) as:
\ba
\lefteqn{
<\psi_{i\alpha}(p'_1) \psi_{k\gamma}(p'_2) | ( S - {\bf 1} ) |
\psi_{j\beta}(p_1) \psi_{l\delta}(p_2) > =}
\nonumber \\
& & = {1 \over Z_\psi^2} \{ <
\bar{u}_\alpha (p'_1) \tilde{G}_t (p_1, p'_1|A)_{ij} u_\beta (p_1) \cdot
\bar{u}_\gamma (p'_2) \tilde{G}_t (p_2, p'_2|A)_{kl} u_\delta (p_2) >_A
\nonumber \\
& & +
< \bar{u}_\gamma (p'_2) \tilde{G}_t (p_1, p'_2|A)_{ki} u_\beta (p_1) \cdot
\bar{u}_\alpha (p'_1) \tilde{G}_t (p_2, p'_1|A)_{il} u_\delta (p_2) >_A \}
{}~.
\ea
$Z_\psi$ is the renormalization constant of the spinor field $\psi$:
$\psi_R (x) = Z_\psi^{-1/2} \psi (x)$.
We have adopted the following relativistic normalization for Dirac spinors:
\be
\bar{u}_\alpha (p) \gamma^\mu u_\beta (p) = 2 p^\mu \delta_{\alpha\beta} ~~~,
{}~~~ \bar{u}_\alpha (p) u_\beta (p) = 2m \delta_{\alpha\beta} ~,
\ee
where $\alpha$ and $\beta$ are spin indices.

The expectation value $< O (A) >_A$ of an arbitrary functional $O (A)$ of
the gluon field $A^\mu$ is now defined as:
\be
< O (A) >_A \equiv {1 \over Z} \displaystyle\int [dA] O (A)
\exp \left[ -{i \over 4} \displaystyle\int d^4 x F_a^{\mu\nu} F_{a \mu\nu}
\right] \det [ i\gamma^\mu D_\mu - m ] ~,
\ee
where $Z$ is the partition function for fermion QCD:
\ba
\lefteqn{
Z \equiv \displaystyle\int [dA] [d\psi] [d\psi^\dagger]
\exp \left[ i \int d^4 x L (\psi ,\psi^\dagger ,A) \right] }
\nonumber \\
& & = \displaystyle\int [dA]
\exp \left[ -{i \over 4} \displaystyle\int d^4 x F_a^{\mu\nu} F_{a \mu\nu}
\right] \det [ i\gamma^\mu D_\mu - m ] ~.
\ea
The determinant in eqs.(3.4) and (3.5) comes from the integration over the
fermion degrees of freedom.

As explained in the previous section, in the high--energy limit we are
considering, $s \to \infty$ with $t,~m^2 \ll s$,
the contribution coming from the second
expectation value in eq.(3.2) ($u$--channel scattering)
is smaller by at least a factor of $s$, when
compared with the first expectation value ($t$--channel scattering),
and hence is negligible.
Therefore, in our limit:
\ba
\lefteqn{
<\psi_{i\alpha}(p'_1) \psi_{k\gamma}(p'_2) | ( S - {\bf 1} ) |
\psi_{j\beta}(p_1) \psi_{l\delta}(p_2) > \simeq }
\nonumber \\
& & \simeq
{1 \over Z_\psi^2}
< \bar{u}_\alpha (p'_1) \tilde{G}_t (p_1, p'_1|A)_{ij} u_\beta (p_1) \cdot
\bar{u}_\gamma (p'_2) \tilde{G}_t (p_2, p'_2|A)_{kl} u_\delta (p_2) >_A ~.
\ea
We must thus evaluate the expectation value:
\be
\tilde{f} (s,t) \equiv
< \bar{u}_\alpha (p'_1) \tilde{G}_t (p_1, p'_1|A)_{ij} u_\beta (p_1) \cdot
\bar{u}_\gamma (p'_2) \tilde{G}_t (p_2, p'_2|A)_{kl} u_\delta (p_2) >_A ~.
\ee
The truncated--connected fermion propagator in the momentum space
is defined as:
\be
\tilde{G}_t (p,p'|A) \equiv \mathop{\lim}_{p^2,p'^2 \to m^2}
{ {\mathaccent 94 p}' - m \over i} \tilde{G} (p,p'|A)
{ {\mathaccent 94 p} - m \over i} ~,
\ee
where $\tilde{G} (p,p'|A)$ is the Fourier transform (see eq.(2.5))
of $G(x,y|A) \equiv <T(\psi(x) \bar{\psi} (y) >_A$. In eq.(3.8) we have
used the notation: ${\mathaccent 94 a} \equiv \gamma^\mu a_\mu$.
As we shall see in the following, in the high--energy limit we are
considering we can make the following replacement in eq.(3.7):
\ba
\lefteqn{
\bar{u}_\alpha (p') \tilde{G}_t (p, p'|A) u_\beta (p) \longrightarrow }
\nonumber \\
& & \longrightarrow {1 \over 2m} \bar{u}_\alpha (p) u_\beta (p) \cdot
\tilde{G}_t^{(scal.)} (p, p'|A) =
\delta_{\alpha\beta} \cdot \tilde{G}_t^{(scal.)} (p, p'|A) ~,
\ea
where we have indicated with $\tilde{G}_t^{(scal.)} (p, p'|A)$
the truncated--connected propagator for a scalar ({\it i.e.} spin--0) quark
in the external gluon field $A^\mu$. Thus we approximately obtain:
\be
\tilde{f} (s,t) \simeq \delta_{\alpha\beta} \delta_{\gamma\delta} \cdot
< \tilde{G}_t^{(scal.)} (p_1, p'_1|A)_{ij}
\tilde{G}_t^{(scal.)} (p_2, p'_2|A)_{kl} >_A ~.
\ee
Let us see how this approximation is justified.
Using the fact that $\bar{u} (p') ({\mathaccent 94 p}' - m) u(p) =
\bar{u} (p') ({\mathaccent 94 p} - m) u(p) = 0$ and the definition (3.8) of
the truncated--connected fermion propagator, one can easily
derive the following perturbative expansion:
\ba
\lefteqn{
\bar{u}_\alpha (p') \tilde{G}_t (p, p'|A) u_\beta (p) =
\bar{u}_\alpha (p')
\left[ -ig \gamma^\mu A_\mu (q_1) \right]_{(1)} u_\beta (p) ~+ }
\nonumber \\
& & +~ \bar{u}_\alpha (p') \left[
(-ig \gamma^\nu A_\nu (q_2)) {i ({\mathaccent 94 p} + {\mathaccent 94 q}_1 + m)
\over (p + q_1)^2 - m^2 + i\varepsilon}
(-ig \gamma^\mu A_\mu (q_1)) \right]_{(2)}
u_\beta (p) + . .
\ea
where we have used the following compact notation:
\ba
\lefteqn{
\left[ F(q_1, \ldots ,q_n) \right]_{(n)} \equiv }
\nonumber \\
& & \equiv \displaystyle\int {d^4 q_1 \over (2\pi)^4} \ldots
\displaystyle\int {d^4 q_n \over (2\pi)^4}
(2\pi)^4 \delta^{(4)} (p' - p - q_1 - \ldots - q_n)
F(q_1, \ldots ,q_n) ~.
\ea
Now, the key--point for the discussion (see, for example,
ref.\cite{Cheng-Wu-book} and references therein) is that,
in the high--energy limit we are considering,
we have $p \simeq p'$ and fermions retain their large
longitudinal momenta during their scattering process. In other words, the
exchanged vector mesons are allowed to carry only transverse momenta, so
that the important phase--space region in the $n$th--order term of the
perturbative expansion (3.11) has the property that ${\mathaccent 94 q}_i$
and $m$ are negligible when compared to ${\mathaccent 94 p}$ in the
numerator (and similarly the terms of the form $q_i q_j$ are negligible
when compared to $p q_i$ in the denominator).
Therefore, apart from a factor $(i)^{n-1}$ coming from the $n-1$ propagators,
the numerator $N (q_1, \ldots ,q_n)$ of the $n$th--order term in (3.11)
can be approximated as:
\ba
\lefteqn{
N (q_1, \ldots ,q_n) \equiv }
\nonumber \\
& & \equiv \bar{u}_\alpha (p') \gamma^{\mu_n} ({\mathaccent 94 p} +
{\mathaccent 94 q}_1 + \ldots + {\mathaccent 94 q}_{n-1} + m)
\gamma^{\mu_{n-1}} \ldots \gamma^{\mu_2} ({\mathaccent 94 p} +
{\mathaccent 94 q}_1 + m) \gamma^{\mu_1} u_\beta (p) \times
\nonumber \\
& & \times (-ig A_{\mu_n} (q_n)) \ldots (-ig A_{\mu_1} (q_1))
\nonumber \\
& & \simeq {1 \over 2m} \bar{u}_\alpha (p) u_\beta (p) \cdot
(2p^{\mu_n}) (2p^{\mu_{n-1}}) \ldots (2p^{\mu_1})
(-ig A_{\mu_n} (q_n)) \ldots (-ig A_{\mu_1} (q_1))
\nonumber \\
& & = \delta_{\alpha\beta} \cdot
(-ig 2p^{\mu_n} A_{\mu_n} (q_n)) (-ig 2p^{\mu_{n-1}} A_{\mu_{n-1}}
(q_{n-1})) \ldots (-ig 2p^{\mu_1} A_{\mu_1} (q_1)) ~.
\ea
Apart from the {\it delta}--function in front, which simply reflects the
fact that fermions retain their helicities during the scattering process
in the high--energy limit, this is
exactly the term we would have expected at the numerator of the $n$th--order
term in the perturbative expansion of the truncated--connected scalar
propagator $\tilde{G}_t^{(scal.)} (p, p'|A)$ in the high--energy limit
(the denominators in (3.11) are already equal to the scalar case!).
In fact, as reported in Table 1 (for scalar QED) and Table 2 (for scalar
QCD), the factors $(-ig 2p^\mu)$ in (3.13) just come from the
quark--quark--gluon vertex $\phi^\dagger \phi A^\mu$ of the scalar theory
in the high--energy limit (when $p \simeq p'$).

\begin{table}[hbt]
\centering
\small
\setlength{\tabcolsep}{1.5pc}
\caption{The quark--photon vertices in scalar QED.}
\vspace{0.3cm}
\label{tab:table1}
\begin{tabular}{rr}
\hline
$Vertex:$ & $Feynman~rule:$ \\
& \\
\hline
& \\
$\phi^\dagger (p_2) \phi (p_1) A^\mu$ & $-i e (p_1 + p_2)^\mu$ \\
& \\
$\phi^\dagger \phi A^\mu A^\nu$ & $i 2 e^2 g^{\mu\nu}$ \\
& \\
\hline
\end{tabular}
\end{table}

\begin{table}[hbt]
\centering
\small
\setlength{\tabcolsep}{1.5pc}
\caption{The quark--gluon vertices in scalar QCD.}
\vspace{0.3cm}
\label{tab:table2}
\begin{tabular}{rr}
\hline
$Vertex:$ & $Feynman~rule:$ \\
& \\
\hline
& \\
$\phi^\dagger_i (p_2) \phi_j (p_1) A^\mu_a$ &
$-i g (p_1 + p_2)^\mu (T_a)_{ij}$ \\
& \\
$\phi^\dagger_i \phi_j A^\mu_a A^\nu_b$ &
$i g^2 g^{\mu\nu} \{ T_a ,T_b \}_{ij}$ \\
& \\
\hline
\end{tabular}
\end{table}

One must observe (see again Table 1 and Table 2)
that the scalar theory has also another type of vertex of
the form $\phi^\dagger \phi A^\mu A^\nu$ (quark--quark--gluon--gluon).
Yet one can easily be convinced that the contributions,
due to these additional scalar vertices, to the $n$th--order
term in the perturbative expansion of the truncated--connected scalar
propagator $\tilde{G}_t^{(scal.)} (p, p'|A)$ in the high--energy limit
are suppressed with respect to the
contribution corresponding to (3.13), {\it i.e.} that coming only from the
quark--quark--gluon couplings. This is essentially due to the fact that the
quark--quark--gluon--gluon vertex does not carry momentum (see Table 1 and
Table 2); moreover, when a vertex of this kind (of order $O(g^2)$) is
inserted in place of two vertices of the quark--quark--gluon kind (each of
them of order $O(g)$), a free fermion propagator (almost on mass--shell)
desappears.

One can also verify directly that the quark--quark--gluon--gluon vertices
do not bring contributions to the quark--quark elastic scattering amplitude
at the leading order in the high--energy limit. For example one can
evaluate the scattering amplitude up to the order $O(g^4)$ in the scalar
theory. The diagrams with only quark--quark--gluon vertices give rise to
the following contribution (see, for example, ref.\cite{Cheng-Wu-book} and
references therein):
\be
M_{(g^4)} \simeq g^2 {2s \over t} [ 1 - \alpha (t) \ln s] \cdot (G_1)_{ij,kl}
+ i g^4 s I(t) \cdot (G_2)_{ij,kl} ~,
\ee
where $G_1$ and $G_2$ are colour factors for the group $SU(N_c)$:
\be
(G_1)_{ij,kl} = (T_a)_{ij} (T_a)_{kl} ~~~ , ~~~
(G_2)_{ij,kl} = (T_a T_b)_{ij} (T_a T_b)_{kl} ~,
\ee
and the functions $\alpha (t)$ and $I(t)$ are defined as:
\be
\alpha (t) = -{N_c g^2 \over 4\pi} t I(t) ~~~ , ~~~
I(t) = \displaystyle\int {d^2 {\bf q}_t \over (2\pi)^2}
{1 \over {\bf q}_t^2 ({\bf \Delta} - {\bf q}_t)^2} ~~~
(t = -{\bf \Delta}^2) ~.
\ee
Instead one finds that the contributions coming from those diagrams with at
least one quark--quark--gluon--gluon vertex are at most of order $O(s^0) =
O(1)$ and so they are suppressed with respect to the terms in (3.14).
Therefore eq.(3.14) is just the quark--quark elastic scattering amplitude
at the leading order in the high--energy limit, up to the perturbative
order $O(g^4)$ in the scalar theory.

Therefore we have found that, in the high--energy limit we are
considering, the replacement (3.9)
is fully justified and we are left with the expression (3.10) for
$\tilde{f} (s,t)$. In the expectation value (3.10) we must consider
$\tilde{G}_t^{(scal.)} (p, p'|A)$ as a functional of the gauge field $A^\mu$
equal to
the truncated--connected propagator for a scalar ({\it i.e.} spin--0) quark
in the external gluon field $A^\mu$.
At this point we can use the results of the previous section, where
$\tilde{G}_t^{(scal.)} (p, p'|A)$ was evaluated in the eikonal
approximation (see eq.(2.7)).
We then proceed exactly as in the previous section, so that we finally find
the following expression for the high--energy quark--quark elastic scattering
amplitude in (fermion) QCD \cite{Nachtmann91}:
\ba
\lefteqn{
M_{fi} = <\psi_{i\alpha}(p'_1) \psi_{k\gamma}(p'_2) | M |
\psi_{j\beta}(p_1) \psi_{l\delta}(p_2) > } \nonumber \\
& & \mathop{\simeq}_{s \to \infty}
-{i \over Z_\psi^2} \cdot \delta_{\alpha\beta} \delta_{\gamma\delta}
\cdot 2s
\displaystyle\int d^2 {\bf z}_t e^{i {\bf q} \cdot {\bf z}_t}
< [ W_1 (z_t) - {\bf 1} ]_{ij} [ W_2 (0) - {\bf 1} ]_{kl} >_A ~.
\ea
The notation is the same as in the previous section.

In a perfectly analogous way one can also derive the high--energy
scattering amplitude in the case of the abelian group $U(1)$ (QED).
The resulting amplitude is equal to eq.(3.17), with the only obvious
difference that now the light--like Wilson lines $W_1$ and $W_2$ are
functionals of the abelian field $A^\mu$ (so they are not matrices).
Thanks to the simple form of the abelian theory (in particular to the
absence of self--interactions among the vector fields), it turns out that
it is possible to explicitly evaluate (at least in the {\it quenched}
approximation) the expectation value of the two
Wilson lines: the details of the calculation are reported in the Appendix
and one finally recovers the well--known result for the eikonal amplitude
of the high--energy scattering in QED \cite{Cheng-Wu}
\cite{Abarbanel-Itzykson} \cite{Jackiw}.

We want to conclude by making a few comments on eq.(3.17) for the case of
QCD. As it was pointed out at the end of the previous section, it is a
singular limit to take the Wilson lines in (3.17) exactly on the
light--cone. It is expected \cite{Verlinde} that a proper regularization of
these singularities will give rise to the $\log s$--dependence of the
amplitude, as obtained by ordinary perturbation theory \cite{Cheng-Wu-book}
\cite{Lipatov} and as confirmed by the experiments on hadron--hadron
scattering processes.

Therefore, the direct evaluation of the expectation value (3.17) is a
highly non--trivial matter, being also strictly connected with the
ultra--violet properties of Wilson--line operators. This last subject was
considered for the first time in ref.\cite{Arefeva80} many years ago.
More recently, in ref.\cite{Korchemsky}, it has been found that there is a
correspondence between high--energy asymptotics in QCD and renormalization
properties of the so--called cross singularities of Wilson lines.
The asymptotic behaviour of the quark--quark scattering amplitude turns out
to be controlled by a $2 \times 2$ matrix of the cross anomalous dimensions
of Wilson lines.

An alternative non--perturbative approach for the calculation of the
expectation value (3.17) has been proposed in ref.\cite{Arefeva94}.
It consists in studying the Regge regime of large energies and fixed
momentum transfers as a special regime of lattice gauge theory on an
asymmetric lattice, with a spacing $a_0$ in the longitudinal direction and
a spacing $a_t$ in the transverse direction. In the limit $a_0 / a_t
\to 0$ one gets a possibility to study correlation functions such as
(3.17) and to obtain the desired $\log s$--dependence of the amplitude.

It would be interesting to investigate about the possibility to use
asymmetric lattice gauge theory to measure directly the amplitude (3.17) by
Monte Carlo simulations.
At the moment, the only non--perturbative numerical estimate of (3.17),
which can be found in the literature, is that of ref.\cite{Dosch} (where
it has been generalized to the case of hadron--hadron scattering): it has
been obtained in the framework of the model of the stochastic vacuum.
It is still an open question whether a more conventional approach based on
numerical simulations of lattice QCD can or cannot be of some help.
Some progress in this direction is expected in the near future.

\bigskip
\noindent {\bf Acknowledgements}
\smallskip

I would like to thank Gabriele Veneziano for many useful discussions and,
first of all, for having prompted my interest in this subject.
I would like to thank also I.Ya. Aref'eva for some useful comments.

\vfill\eject

\renewcommand{\thesection}{A}
\renewcommand{\thesubsection}{A.\arabic{subsection}}

\pagebreak[3]
\setcounter{section}{1}
\setcounter{equation}{0}
\setcounter{subsection}{0}
\setcounter{footnote}{0}

\begin{flushleft}
{\bf Appendix: The abelian case.}
\end{flushleft}

\noindent
In this appendix we shall discuss the abelian case (See also
refs.\cite{Korchemsky} and \cite{Arefeva94}).
The fermion--fermion (or scalar--scalar) electro--magnetic scattering
amplitude, in the high--energy limit $s \to \infty$ and $t \ll s$, can be
derived following exactly the same procedure used in Sect. 2 and Sect. 3
for the non--abelian case.
The resulting amplitude (for the fermion case)
is formally identical to eq.(3.17), with the only obvious
difference that now the Wilson light--like lines $W_1$ and $W_2$ are
functions of the abelian field $A^\mu$ (so they are not matrices):
\ba
\lefteqn{
M_{fi} = <\psi_{\alpha}(p'_1) \psi_{\gamma}(p'_2) | M |
\psi_{\beta}(p_1) \psi_{\delta}(p_2) > } \nonumber \\
& & \mathop{\simeq}_{s \to \infty}
-{i \over Z_\psi^2} \cdot \delta_{\alpha\beta} \delta_{\gamma\delta}
\cdot 2s
\displaystyle\int d^2 {\bf z}_t e^{i {\bf q} \cdot {\bf z}_t}
< [ W_1 (z_t) - 1 ] [ W_2 (0) - 1 ] >_A ~.
\ea
The electro--magnetic light--like Wilson lines $W_1$ and $W_2$ are defined
as in (2.33):
\ba
W_1 (z_t) \equiv W_1 (+\infty , 0, {\bf z}_t) &=&
\exp \left[ -ie \displaystyle\int_{-\infty}^{+\infty}
A_\mu (z_t + p_1 \tau) p_1^\mu d\tau \right] ~,
\nonumber \\
W_2 (0) \equiv W_2 (0, +\infty , {\bf 0}_t) &=&
\exp \left[ -ie \displaystyle\int_{-\infty}^{+\infty}
A_\mu (p_2 \tau) p_2^\mu d\tau \right] ~,
\ea
where $z_t = (0,0,{\bf z}_t)$ and $e$ is the electric coupling--constant
(electric charge).
Thanks to the simple form of the abelian theory (in particular to the
absence of self--interactions among the vector fields), it turns out that
it is possible to explicitly evaluate the expectation value of the two
Wilson lines, thus finally recovering the well--known result for the eikonal
amplitude of the high--energy scattering in QED (see refs. \cite{Cheng-Wu},
\cite{Abarbanel-Itzykson} and \cite{Jackiw}).
Let us see the details of the calculation.

We shall evaluate the expectation value $< W_1 (z_t) W_2 (0) >_A$ in the
so--called {\it quenched} approximation, where vacuum polarization effects,
arising from the presence of loops of dynamical fermions, are neglected.
This amounts to setting $\det (K[A]) = 1$, where $K[A] = i\gamma^\mu D_\mu
- m$ is the fermion matrix.
We shall also put $Z_\psi \simeq 1$.
Thus we can write that:
\be
< W_1 (z_t) W_2 (0) >_A \simeq
{1 \over Z} \int [dA_\mu] e^{iS_A} W_1 (z_t) W_2 (0) ~,
\ee
where $S_A = -{1 \over 4} \int d^4 x F_{\mu\nu} F^{\mu\nu}$ is the action
of the electro--magnetic field and $Z = \int [dA_\mu] e^{iS_A}$ is the
pure--gauge partition function.
We then add to the pure--gauge Lagrangian $L_A = -{1 \over 4} F_{\mu\nu}
F^{\mu\nu}$ a {\it gauge--fixing} term $L_{GF} = -{1 \over 2\alpha}
(\partial^\mu A_\mu)^2$ ({\it covariant} or {\it Lorentz} gauge). The
expectation value (A.3) becomes, denoting $L_0^F = L_A + L_{GF}$:
\ba
\lefteqn{
< W_1 (z_t) W_2 (0) >_A \simeq
{1 \over Z'} \int [dA] \times } \nonumber \\
& & \times \exp \left[ i \left( \displaystyle\int d^4 x L_0^F
-e \displaystyle\int_{-\infty}^{+\infty}
A_\mu (z_t + p_1 \tau) p_1^\mu d\tau
-e \displaystyle\int_{-\infty}^{+\infty}
A_\mu (p_2 \tau) p_2^\mu d\tau \right) \right] ~,
\ea
with $Z' = \int [dA] \exp(i\int d^4 x L_0^F)$.
It will be helpful, for the computation, to write the sum of the two line
integrals as a 4--volume integral of the form:
\be
-e \displaystyle\int_{-\infty}^{+\infty}
A_\mu (z_t + p_1 \tau) p_1^\mu d\tau
-e \displaystyle\int_{-\infty}^{+\infty}
A_\mu (p_2 \tau) p_2^\mu d\tau
= \int d^4 x A_\mu (x) J^\mu (x) ~,
\ee
where $J^\mu (x)$ is a 4--vector source defined as:
\be
J^\mu (x) = J^\mu_{(+)} (x) + J^\mu_{(-)} (x) \nonumber \\
= -e [ \epsilon^\mu_{(+)} \delta (x_-) \delta ({\bf x}_t - {\bf z}_t)
+ \epsilon^\mu_{(-)} \delta (x_+) \delta ({\bf x}_t) ] ~,
\ee
with: $\epsilon^\mu_{(+)} = (1,1,0,0)$ and $\epsilon^\mu_{(-)} =
(1,-1,0,0)$.
Therefore the exponent in eq.(A.4) can be written as:
\be
i \int d^4 x ( L_0^F + A_\mu J^\mu ) ~.
\ee
After integration by parts, the integral of $L_0^F$ can be written as an
integral of a quadratic form of the electro--magnetic field $A^\mu$:
\be
\int d^4 x L_0^F = \displaystyle\int d^4 x \left[ -{1 \over 4} F_{\mu\nu}
F^{\mu\nu} - {1 \over 2\alpha} (\partial^\mu A_\mu)^2 \right]
= \displaystyle\int d^4 x \left[ -{1 \over 2} A_\mu K^{\mu\nu} A_\nu
\right] ~,
\ee
and the (invertible!) matrix $K_{\mu\nu}$ is so defined:
\be
K_{\mu\nu} = -g_{\mu\nu} \partial^2 + \left( 1 - {1 \over \alpha} \right)
\partial_\mu \partial_\nu ~.
\ee
The functional integral
\ba
\lefteqn{
Z_0 [J] \equiv \int [dA] \exp \left[ i\int d^4 x ( L_0^F + A_\mu J^\mu )
\right] }
\nonumber \\
& & = \int [dA] \exp \left[ i\int d^4 x \left( -{1 \over 2} A_\mu K^{\mu\nu}
A_\nu + A_\mu J^\mu \right) \right]
\ea
can be evaluated with standard methods (completing the quadratic form in
the exponent). One thus obtains:
\be
Z_0 [J] = Z' \cdot \exp \left[ {i \over 2} \int d^4 x \int d^4 y
J^\mu (x) D_{\mu\nu} (x-y) J^\nu (y) \right] ~,
\ee
where $Z' = Z_0 [J=0] = \int [dA] \exp(i\int d^4 x L_0^F)$ and
$D_{\mu\nu}$ is the inverse of the operator $K_{\mu\nu}$:
\be
\int d^4 z  K_{\mu\lambda} (x-z) g^{\lambda\rho} D_{\rho\nu} (z-y)
= g_{\mu\nu} \delta^{(4)} (x-y) ~.
\ee
In other words, $D_{\mu\nu}$ is the free photon propagator and is equal to:
\be
D_{\mu\nu} (x-y) = \displaystyle\int {d^4 k \over (2\pi)^4}
{e^{-ik(x-y)} \over k^2 +i\varepsilon}
\left( g_{\mu\nu} - (1 - \alpha){k_\mu k_\nu \over k^2} \right) ~.
\ee
In the following we shall choose the gauge--fixing parameter $\alpha$ equal
to $1$. From eqs.(A.4), (A.11) and (A.13), we derive the following expression
for the expectation value of the two Wilson lines (A.2):
\ba
\lefteqn{
< W_1 (z_t) W_2 (0) >_A \simeq
{1 \over Z'} \int [dA] e^{i\int d^4 x ( L_0^F + A_\mu J^\mu ) } }
\nonumber \\
& & = Z_0 [J] / Z' = \exp \left[ {i \over 2} \int d^4 x \int d^4 y
J^\mu (x) J_\mu (y) \displaystyle\int {d^4 k \over (2\pi)^4}
{e^{-ik(x-y)} \over k^2 +i\varepsilon} \right] ~.
\ea
Proceeding in exactly the same way, we can also derive
(always in the {\it quenched} approximation) the expectation values
of the single Wilson lines $W_1 (z_t)$ and $W_2 (0)$:
\ba
< W_1 (z_t) >_A &\simeq& \exp \left[ {i \over 2} \int d^4 x \int d^4 y
J_{(+)}^\mu (x) J_{(+)\mu} (y)
\displaystyle\int {d^4 k \over (2\pi)^4}
{e^{-ik(x-y)} \over k^2 +i\varepsilon}
\right] ~, \nonumber \\
< W_2 (0) >_A &\simeq& \exp \left[ {i \over 2} \int d^4 x \int d^4 y
J_{(-)}^\mu (x) J_{(-)\mu} (y)
\displaystyle\int {d^4 k \over (2\pi)^4}
{e^{-ik(x-y)} \over k^2 +i\varepsilon}
\right] ~,
\ea
where the 4--vector sources $J_{(+)}^\mu$ and $J_{(-)}^\mu$ are defined in
eq.(A.6). Using the fact that: $\epsilon_{(+)}^\mu \epsilon_{(+)\mu} =
\epsilon_{(-)}^\mu \epsilon_{(-)\mu} = 0$, one finds that:
\be
< W_1 (z_t) >_A = < W_2 (0) >_A \simeq 1 ~.
\ee
Finally, using the explicit form (A.6) of the 4--vector source $J^\mu (x)$ to
evaluate the double integral in eq.(A.14), one finds the
well--known eikonal form of the high--energy electro--magnetic scattering
amplitude \cite{Cheng-Wu} \cite{Abarbanel-Itzykson} \cite{Jackiw}):
\ba
\lefteqn{
M_{fi} \simeq -i \cdot \delta_{\alpha\beta} \delta_{\gamma\delta}
\cdot 2s
\displaystyle\int d^2 {\bf z}_t e^{i {\bf q} \cdot {\bf z}_t}
< [ W_1 (z_t) - 1 ] \cdot
[ W_2 (0) - 1 ] >_A }
\nonumber \\
& & \simeq -i \cdot \delta_{\alpha\beta} \delta_{\gamma\delta}
\cdot 2s
\displaystyle\int d^2 {\bf z}_t e^{i {\bf q} \cdot {\bf z}_t}
[ < W_1 (z_t) W_2 (0) >_A - 1 ]
\nonumber \\
& & \simeq -i \cdot \delta_{\alpha\beta} \delta_{\gamma\delta}
\cdot 2s
\displaystyle\int d^2 {\bf z}_t e^{i {\bf q} \cdot {\bf z}_t}
\left( \exp \left[ -i e^2 \displaystyle\int {d^2 {\bf k}_t \over (2\pi)^2}
{e^{-i{\bf k}_t \cdot {\bf z}_t} \over {\bf k}_t^2 - i\varepsilon} \right]
- 1 \right) ~.
\ea
We could have also proceeded in a slightly different way, by observing that
the free photon propagator has a rather simple expression in the coordinate
space. Explicitly, in the Lorentz gauge with $\alpha = 1$:
\be
D_{\mu\nu} (x) = g_{\mu\nu} {i \over 4\pi^2 (x^2-i\varepsilon)} ~.
\ee
After inserting this expression into eq.(A.11) and integrating with respect
to $x$ and $y$, we are left with the following expression for the
scattering amplitude:
\be
M_{fi} \simeq -i \cdot \delta_{\alpha\beta} \delta_{\gamma\delta}
\cdot 2s
\displaystyle\int d^2 {\bf z}_t e^{i {\bf q} \cdot {\bf z}_t}
\left( \exp \left[ {e^2 \over 8\pi^2} \displaystyle\int {d\alpha d\beta \over
\alpha \beta + {\bf z}_t^2 + i\varepsilon} \right]
- 1 \right) ~.
\ee
This expression is, of course, perfectly equivalent to that in eq.(A.17), as
one can easily verify by explicitly evaluating the two integrals in the
exponents. In fact, the two exponentials in eq.(A.17) and eq.(A.19)
turn out to be equal to:
\be
e^{-ie^2 \Lambda} \cdot \exp \left( i {e^2 \over 2\pi} \ln |{\bf z}| \right) ~,
\ee
where $\Lambda$ is an infinite constant phase and is therefore physically
unobservable. The origin of this infinite constant phase resides in the
well--known fact that the fermion--fermion scattering amplitude in QED has
infrared divergences, due to the emission of low--energy massless vector
mesons. The traditional way to handle these infrared divergences is to
introduce an infrared cutoff in the form of a vector meson mass $\mu$.
In this way the integral over ${\bf k}_t$ in the exponent of eq.(A.17) is
substituted by the following expression:
\be
\displaystyle\int {d^2 {\bf k}_t \over (2\pi)^2}
{e^{-i{\bf k}_t \cdot {\bf z}_t} \over {\bf k}_t^2 + \mu^2} \equiv
{1 \over 2\pi} K_0 (\mu |{\bf z}_t|) ~,
\ee
where $K_0$ is the modified Bessel function. In the limit of small $\mu$
this last expression can be replaced by:
\be
{1 \over 2\pi} K_0 (\mu |{\bf z}|) \mathop{\simeq}_{\mu \to 0}
-{1 \over 2\pi} \ln \left( {1 \over 2} e^\gamma \mu |{\bf z}_t| \right) ~.
\ee
Absorbing ${1 \over 2} e^\gamma$ in $\mu$ and putting
$\Lambda = -(1/2\pi) \ln ({1 \over 2} e^\gamma \mu)$ (so that $\Lambda
\to \infty$ when $\mu \to 0$), we just obtain the expression (A.20) for
the exponential in eq.(A.17).
The integral in (A.17) with respect to the transverse coordinate ${\bf z}_t$
can be explicitly performed and one finally gets:
\be
M_{fi} \simeq \delta_{\alpha\beta} \delta_{\gamma\delta} \cdot 2s \cdot
{ \Gamma (1 + i \alpha) \over 4 \pi i \mu^2 \Gamma (-i \alpha) }
\displaystyle\left( { 4 \mu^2 \over -t } \right)^{1 + i \alpha} ~,
\ee
where: $\alpha = e^2 / 4 \pi$. This is just the standard eikonal formula
for the fermion--fermion electro--magnetic scattering amplitude.

\vfill\eject

{\renewcommand{\Large}{\normalsize}
}

\vfill\eject

\end{document}